# Realization of Universal Quantum Gates with Spin-Qudits in Colloidal Quantum Dots

Fabrizio Moro,[*] Alistair J. Fielding, Lyudmila Turyanska, and Amalia Patanè

Dr. F. Moro
Department of Materials Science, University of Milano-Bicocca, 20125, Milano, Italy
School of Physics and Astronomy, University of Nottingham, NG7 2RD, Nottingham, UK
fabrizio.moro@unmib.it

Dr. A. J. Fielding
School of Pharmacy and Biomolecular Sciences, Liverpool John Moores University, Byrom Street, L3 3AF, Liverpool, UK

Dr L. Turyanska
School of Chemistry, University of Lincoln, LN6 7DL, Lincoln,UK
School of Physics and Astronomy, University of Nottingham, NG7 2RD, Nottingham, UK

Prof. A. Patanè
School of Physics and Astronomy, University of Nottingham, NG7 2RD, Nottingham, UK

## Abstract

We exploit hyperfine interactions in a single Mn-ion confined in a quantum dot (QD) to create a *qudit*, *i.e.* a multi-level quantum-bit system, with well-defined, addressable and robust set of spin states for the realization of universal quantum gates. We generate and probe an arbitrary superposition of states between selected hyperfine energy level pairs by using electron double resonance detected nuclear magnetic resonance (EDNMR). This enables the observation of Rabi oscillations and the experimental realization of NOT and $\sqrt{\text{SWAP}}$ universal quantum gates that are robust against decoherence. Our protocol for cyclical preparation, manipulation and read-out of logic gates offers opportunities for integration of *qudits* in scalable quantum circuit architectures beyond solid state electron spin *qubits*.

# 1. Introduction

Quantum computers rely on a quantum information unit, *i.e.* the *qubit*, which exists in a 0 or 1 state, as well as in any superposition of these two states.[1],[2] Requirements for the implementation of quantum computation involve several challenges, including the suppression of decoherence phenomena originating from the interactions of *qubits* with their environment, while enabling the manipulation of an ensemble of weakly interacting *qubits*. This conundrum can be partially overcome by the realization of a multi-level quantum-bit system , *i.e.* a *qudit*,[3] which offers several advantages compared to a *qubit*, such as a multi-dimensional Hilbert space ($d > 2$) for encoding several bits per unit, reduced number of units and hardware size, increased scalability and robustness against noise and error rates.[3],[4]

Potential building blocks for *qudits* are intensively investigated and include electron spins in quantum dots (QDs),[5] in molecular magnets,[6],[7] and in solids (*e.g.* NV-centres in diamond[8] and Si[9]), rotational and vibrational states of molecules,[10] harmonic oscillator states in superconducting cavities[11] and circuits,[12] and hyperfine levels of alkali atoms.[13] Among various quantum systems, of particular interest are transition metal (*e.g.* Mn, Fe) and rare earth (*e.g.* Gd, Tb) ions embedded in solid state,[14] in molecular systems[7, 15] and QDs.[16],[17],[18] The hyperfine interaction between electrons on 3d or 4f orbitals and the nuclear spin creates a multi-level system, and hence a *qudit*, which could encode and store information.[19] Although this type of *qudit* has been studied, its experimental implementation in universal quantum gates has not yet been achieved in low dimensional semiconductor systems. This further development may facilitate the integration of *qudits* in quantum computer architectures with large storage capabilities and scalability.

Amongst various semiconducting systems, a promising way of controlling and exploiting confined electron spins is through colloidal PbS QDs. These nanocrystals consist of semiconductor materials (*e.g.* PbS) surrounded by capping ligands. The controlled incorporation of magnetic impurities (*e.g.* Mn) with concentration down to a single impurity per QD[17c],[20] and flexibility in the manipulation of the QD surface and environment enable the minimization of the major sources of Mn spin *qubit* decoherence (*i.e.* Mn-Mn spin interactions, protons of the QD surface capping ligands and nuclear spins) and to observe long quantum coherence near room temperature.[18, 21],[17a] Furthermore, these narrow band gap nanocrystals[22],[23] offer opportunities for optical control of exciton-Mn entangled states by ultrafast optical pulses[24] and integration with other low-dimensional semiconductors, such as graphene, to construct new hybrid functional devices.[25]

In this work, we realize experimentally a *qudit* and quantum gates in individual manganese ions confined in colloidal PbS QDs (**Figure 1a**) with electron-nuclear spin states $S = 5/2$ and $I = 5/2$ (**Figure 1b**) by using electron double resonance detected nuclear magnetic resonance (EDNMR). This technique combines electron spin resonance (ESR) and nuclear magnetic resonance (NMR). EDNMR uses a high-turning angle microwave pulse to drive spin forbidden transitions with selection rules $\Delta m_s = \pm 1$, $\Delta m_I \neq 0$ within the electron-nuclear energy manifold, therefore allowing the nuclear transitions to be probed indirectly. Thus, we use EDNMR to resolve and manipulate allowed and forbidden quantum transitions between electron-nuclear spin states of $Mn^{2+}$ ions in QDs. We demonstrate the generation of arbitrary superposition of states and Rabi oscillations, implementing NOT and $\sqrt{SWAP}$ universal quantum gates[26] (**Figure 1c**). The coherent manipulation of multiple levels is realized by dynamically decoupling Mn-spins from dipolar interactions with surrounding nuclear spins. This approach could be applied to other quantum systems and represents an important step towards the implementation of spin-*qudits* in downsized and scalable quantum computing architectures.

## 2. Results and discussion

### 2.1 EDNMR spectra

**Figure 1b** shows the full echo detected field swept spectrum of the QDs. The six-line pattern is characteristic for the isotropic hyperfine splitting $A_{iso}$ of isolated $Mn^{2+}$ ($S = 5/2$ and $I = 5/2$). The lines originate from the zero quantum transitions, ZQ (*i.e.* allowed ESR transitions, $\Delta m_s = \pm 1$, $\Delta m_I = 0$) within the $m_s = |\pm 1/2\rangle$ spin multiplet. The broad wings beside the central narrow hyperfine lines are ascribed to the contribution of forth order zero field splitting (ZFS) terms due to Mn ions that experience distortions of the PbS cubic lattice near the QD surface.[18] The $m_s = |+ 1/2\rangle$ and $m_s = |- 1/2\rangle$ states are chosen for the definition of the *qubit/qudit* states because their hyperfine lines are isotropic and are not broadened by the orientation distribution and strain of the ZFS parameters. Thus, the hyperfine lines of the $m_s = |+ 1/2\rangle$ and $m_s = |- 1/2\rangle$ states can be spectroscopically resolved and manipulated with microwave pulses, as shown below.

The electron spin resonance absorption spectrum in **Figure 1b** can be described by the spin-Hamiltonian:[27]

$$\hat{H} = g_e \mu_B \mathbf{B}\hat{\mathbf{S}} - g_N \mu_N \mathbf{B}\hat{\mathbf{I}} + \sum_{-4}^{4} B_4^q \hat{O}_4^q(S) + hA_{iso}\hat{\mathbf{S}}\hat{\mathbf{I}}, \qquad (1)$$

where $g_e$ and $g_N$ are the Landé g-factor for electron and nuclear spin, respectively, $\mu_B$ and $\mu_N$ are the Bohr magneton for electron and nuclear spin, respectively, **B** is the magnetic field vector, $h$ is the Planck constant, $B_4^q$ are the ZFS fourth order coefficients of the extended Stevens operators $\hat{O}_k^q$, and $S$ and $I$ are the electron and nuclear spin quantum numbers, respectively. The hyperfine term leads to a positive energy shift, $+hA_{iso}/4$, of the energy levels associated to $m_s$ and $m_I$ states with the same sign and to a negative energy shift, $-hA_{iso}/4$, of the energy levels associated to $m_s$ and $m_I$ states with opposite sign. In combination with the nuclear Zeeman term, this leads to a different energy spacing of the levels within the multiplets with $m_s = |- 1/2\rangle$ and $m_s = |+ 1/2\rangle$, and hence to the observation of the 6 hyperfine lines for $I = 5/2$ (**Figure 1b** and **Figure 2a**).

The sextet is centered at $g_e = 2.0025$, close to the free electron $g_e$-value, with an isotropic hyperfine constant $|A_{iso}| \sim 267$ MHz ($\sim 9.3$ mT), **Figure 2a**. An accurate estimate of the ZFS contribution is hindered by the linewidth broadening due to strain and/or disorder. This causes unresolved hyperfine lines originating from the states with $m_s \neq 1/2$. However, the wings beside the central hyperfine suggest that ZFS $\gg A_{iso}$, and that for a selected central hyperfine line the overlapping contribution of hyperfine transitions involving states with $m_s \neq 1/2$ is avoided.

**Figure 2b** shows the EDNMR spectrum acquired by scanning the electron double resonance (ELDOR) pulse frequency, $\nu_{ELDOR}$, while recording a free induction decay (FID) or primary spin echo signal on the observer channel.[28] The EDNMR spectrum shows: the full quenching of the FID for $\Delta\nu = \nu_{ELDOR} - \nu_{obs} = 0$, where $\nu_{obs}$ is the observation frequency; two doublets centered at $\Delta\nu \sim + 139$ MHz and $\sim + 279$ MHz; one doublet centered at $\Delta\nu \sim - 139$ MHz; and an additional peak at $\Delta\nu \sim 51$ MHz.

For EDNMR experiments performed at the ESR transition $m_S = |-1/2\rangle \rightarrow |+1/2\rangle$, the EDNMR spectrum results in two single quantum transitions (SQ' and SQ" with $\Delta m_I = \pm 1$, centered at $\Delta\nu = \pm |A_{iso}|/2$ and separated by $2\nu_I$, where $\nu_I = g_N \mu_N \mathbf{B}/h$ is the nuclear spin Larmor frequency), and two double quantum transitions (DQ' and DQ" with $\Delta m_I = \pm 2$, centered at $\Delta\nu = |A_{iso}|$ and separated by $4\nu_I$) symmetrically distributed around the ESR transition. In addition, a DQ transition connecting the $m_I$ states $|+3/2\rangle$ and $|-1/2\rangle$ and overlapping with the SQ' transition could be observed. For a given EDNMR pulse length, the contribution of the DQ transitions is much smaller than that of the SQ transitions. Also, only transitions that share a level with the central ZQ transition (*i.e.* $m_I = |-1/2\rangle$) can be observed in EDNMR.

Our assignment of the ELDOR lines at Δν = 0 to the ZQ transitions and the doublets at Δν ~ ± 139 and + 279 MHz to the SQ and DQ transitions is confirmed by the measured resonance frequencies, which are close to the isotropic $|A_{iso}|/2$ (~ ± 134 MHz) and $|A_{iso}|$ (~ + 267 MHz) NMR transitions between the Mn hyperfine spin states in the strong coupling regime ($|A_{iso}| \gg 2\nu_I$ ($\nu_I$ ~ 12.8 MHz for $^{55}$Mn and $B_0$ ~ 1213 mT). However, the separation between the peaks in the SQ ( ~ 43 MHz) and DQ ( ~ 84 MHz) doublets[29],[30] are larger than the expected values, $2\nu_I$ and $4\nu_I$, respectively, revealing the contribution of second order hyperfine interactions:[29-30]

$$2\nu_I = 2g_N\mu_N B/h + \frac{A^2}{\nu_0}[S(S+1) - m_s^2] - \frac{A^3}{\nu_0^2}[S(S+1) - m_s^2] - O(A^3). \quad (2)$$

For $B$ = 1213 mT, $A \equiv A_{iso}$ = 267 MHz, $\nu_0$ = 34 GHz, $S$ = 5/2 and $m_s$ = 1/2, we find that $2\nu_I$ ~ 43 MHz and $4\nu_I$ ~ 86 MHz, in agreement with the experimental data. Finally, the peak at Δν ~ 51 MHz is assigned to hydrogen-atoms on the QD capping ligands weakly coupled to Mn spins.[21]

**2.2 Multi-level manipulations**

The EDNMR data demonstrate the existence of distinguishable and addressable multiple energy levels in isolated Mn$^{2+}$ ions arising from hyperfine interactions. EDNMR performed on the ZQ transition |-1/2 -1/2⟩ → |+1/2 -1/2⟩ enables to select a subset of eight energy levels connected by seven NMR transitions (**Figure 2b**). As discussed above, the NMR transitions |-1/2 +3/2⟩ → |+1/2 -1/2⟩ and |-1/2 -1/2⟩ → |+1/2 +1/2⟩ coincide, thus leaving six transitions connecting seven energy levels, *i.e.* |-1/2 -5/2⟩, |-1/2 -3/2⟩, |-1/2 -1/2⟩, |-1/2 +1/2⟩, |+1/2 -1/2⟩, |+1/2 -3/2⟩, |+1/2 -5/2⟩ that can be unequivocally chosen. These are linked by a combination of addressable EDNMR and NMR transitions to form a universal basis set corresponding to a *qudit* with dimension *d* = 7. If we consider only EDNMR transitions, the universal set is realized by the three states involved in the ZQ transition and any of the SQ or DQ transitions. Thus the *qudit* dimension is reduced to $d = 3$.

The next step towards the realization of universal quantum gates is the coherent manipulation of the energy level population across two or more energy levels. We perform nutation-EDNMR experiments on the allowed ESR transition $m_S = |-1/2\rangle \rightarrow |+1/2\rangle$, to demonstrate single-*qubit* operations, *i.e.* two-levels manipulation, by tuning the ELDOR nutation frequency to the observation frequency, *i.e.* $\nu_{ELDOR} = \nu_{obs}$ (inset in **Figure 3a**). The FID is detected as a function of the ELDOR pulse duration and reveals Rabi oscillations with a dominant nutation frequency $\Omega_R \sim 9$ MHz at maximum microwave field, $B_1$ (**Figure 3a**). The decay of the oscillations during the nutation pulse increments and distribution of nutation frequencies are ascribed to the population transfer between energetically close $m_I$ states, providing a "leakage" pathway during the coherent evolution of electron and nuclear spins as well as to the inhomogeneous microwave field $B_1$.[31],[33] We also note that the Rabi oscillations for the ZQ transitions $|-1/2\ -1/2\rangle \rightarrow |+1/2\ -1/2\rangle$ and $|-1/2\ -3/2\rangle \rightarrow |+1/2\ -3/2\rangle$ do not show significant differences thus confirming that magnetic anisotropy does not affect the transition frequencies (**Figure S1**).

The microwave power dependence of the nutation-EDNMR experiments and analysis of their Fast Fourier Transform (FFT) show a distribution of nutation frequencies with a dominant contribution that scales linearly with the microwave field $B_1$ for the ZQ transition (**Figure 3a** and **Figure S2**), supporting the observation of coherent oscillations and the generation of arbitrary superpositions of states.

We implement two-*qubit* operations on the subset of energy levels involving the ZQ and SQ transitions and sharing an energy level, by tuning $\nu_{ELDOR}$ to the SQ' transition $|-1/2\ -3/2\rangle \rightarrow |+1/2\ -1/2\rangle$, and $\nu_{obs}$ to the ZQ transition $|-1/2\ -1/2\rangle \rightarrow |+1/2\ -1/2\rangle$. In this way a subset of three energy levels can be selected by EDNMR and the $^{55}$Mn ions in the QDs can be treated

as a system with an effective electron spin $S = 1/2$ interacting with a nucleus with $I = 1/2$ (**Figure 1c**). As the ELDOR pulse duration is increased, the population difference between the SQ levels |-1/2 -3/2⟩ and |+1/2 -1/2⟩ begins to oscillate periodically changing the population difference between the ZQ transition levels |-1/2 -1/2⟩ and |+1/2 -1/2⟩. Thus, the Rabi oscillations between the SQ' energy levels are indirectly probed by the oscillation of the echo signal in the detection channel. The results are reported in **Figure 3b** along with those obtained on the subset of energy levels involving the ZQ and DQ' transitions. The comparison in **Figure 3b** shows that the Rabi frequency for the DQ' transition is smaller than that observed for the SQ' and ZQ transitions. This is because the Rabi frequency is proportional to the transition probabilities of the energy levels involved.[28]

**2.3 Quantum coherence**

Quantum coherence was studied using the primary spin echo as well as by a multi-pulse Carr-Purcell-Meiboom-Gill (CPMG) pulse sequence[28] (inset of **Figure 4**) in an ensemble of QDs. In the primary echo experiments the echo signal is measured as function of the π/2 and π pulses with interpulse distance 2τ (**Figure 2a**). In the CPMG pulse sequence, a π/2 pulse is applied along the *x*-axis in the ESR rotating frame followed by a series of π pulses applied along the *y*-axis at times 2n + 1 for n = 0, 1, ...N, yielding multiple echoes at times 2n + 2, which are recorded as function of 2τ. **Figure 4** shows the experimental results and their analysis. The fit of the primary echo data to an exponential decay function gives a decay time constant $T_M \sim 1$ μs. The fit of the CPMG echo signal to a bi-exponential decay function yields a long spin-spin relaxation time constant, $T'_{CPMG} \sim 8$ μs, and a short time constant, $T''_{CPMG} \sim 1$ μs. We suggest that $T'_{CPMG} > T_M$ is due to the train of refocusing $\pi_y$ CPMG pulses that enable effective dynamic decoupling of the observer spin from spectral diffusion effects induced by random fluctuations of the surrounding nuclear spins.[32] Also, we ascribe $T''_{CPMG}$ to a contribution of electron-electron dipolar coupling of the nearby Mn spins. From the analysis of the amplitude of the bi-exponential decay fitting, we estimate that these contribute about 50%. The high frequency electron-electron dipolar coupling is not suppressed by the CPMG sequence because the latter enables the suppression of decoherence sources with frequencies lower that $1/\tau = 5$ MHz.[33],[34]

Thus, individual $Mn^{2+}$-ions confined in QDs possess properties that fulfill key requirements for their exploitation as *qudits*:

(i) A long quantum coherence time ($T'_{CPMG} \sim 8$ μs) in an ensemble of QDs, which is much longer than the π/2 microwave pulse (~ 24 ns) used for the *qudit* manipulation.

(ii) The large hyperfine coupling in a $Mn^{2+}$ ion induces well-defined and distinguishable EDNMR transitions $|-1/2\ -1/2\rangle \rightarrow |+1/2\ -1/2\rangle$, $|-1/2\ -3/2\rangle \rightarrow |+1/2\ -1/2\rangle$ and $|-1/2\ -5/2\rangle \rightarrow |+1/2\ -1/2\rangle$ occurring in the MHz energy range, which is within the resonator bandwidth and cannot be achieved either with exchange-only spin *qubits* in double QDs (*e.g. J* ~ 80 μeV ~ 20 GHz)[1] or with transitions originating from fine interactions in *qudits* based on single-ion molecular magnets (*e.g.* ~ 2 GHz in $Gd^{3+}$).[6c]

(iii) An effective manipulation of the energy level population can be achieved by using short (20 - 200 ns) Rabi turning pulses. This is possible because transition probabilities between electron–nuclear levels are intrinsically larger for $I = 5/2$, and the power required to pump/rotate these transitions is smaller than that required to pump/rotate low spin nuclei (*i.e. I* = 1/2), (**Figure S3**, supplementary information).[35]

**2.4 Realization of universal quantum gates**

We now discuss these findings within the quantum information processing framework. The Rabi oscillations of the ZQ transition represent an exact logic NOT gate: the ZQ transition connects two levels with each of them sharing an energy level with either the SQ' or the DQ' transitions. When the ELDOR pulse is tuned to the SQ' transition, it transforms the $|-1/2\ +1/2\rangle$ state into a superposition of the $|-1/2\ +1/2\rangle$ and $|+1/2\ -1/2\rangle$ states. This simultaneously and coherently flips both electron and nuclear spins (**Figure 1b**). In particular, when the ELDOR pulse saturates the excited transition, *i.e.* for an ELDOR pulse duration of ~ 80 ns, a SWAP gate is realized; in contrast, when the ELDOR pulse duration approaches ~ 40 ns, an half SWAP gate, that is a $\sqrt{SWAP}$ gate, is realized. We stress that the realization of the entangled state ($|01>+i|10>)/\sqrt{2}$ requires control over the pulse phase, which is not examined in this study.

EDNMR enables the execution of cyclical logical operations involving preparation, manipulation and read-out (**Figure 5**). The initialization of an initial quantum state $|i>$ of the Mn hyperfine structure is achieved by application of a magnetic field that matches the resonance condition with the detection frequency. Furthermore, for a positive and sufficiently large ZFS the $m_s = |\pm 1/2>$ states represent the ground state at conventional X-band (~ 10 GHz) and Q-band (~ 34 GHz) frequencies. Thus, these states could be simply prepared by cooling. To fully realize the potential of isolated Mn-ions as *qudits*, the mono-nuclear Mn complexes with positive and large ZFS (*i.e.* $B_2^0 > 10^3$ MHz)[36] could be used as seeds for the growth of doped-QDs.[37] Alternatively, the QD surface could be engineered in such a way that the embedded Mn ions nearby or at the surface[38] experience a sufficiently large ZFS due to surface disorder and lower symmetry.

The manipulation process initiates with the ELDOR pulse tuned to a particular frequency that matches the energy separation between the initial ($|i>$) and final ($|f>$) state. Reading out the population of $|f>$ is achieved by the detection of coherences between $|i>$ and an auxiliary state $|aux>$.

Based on our findings, the gating time ($T_g$ ~ 80 ns) required to perform a cyclical $\sqrt{SWAP}$ operation by exploiting the SQ' transition is shorter than that involving a DQ' transition ($T_g$ ~ 250 ns). This is due to the lower microwave power available at the DQ' frequency. For $T'_{CPMG}$ ~ 10 μs, it is possible to realize about $N$ ~ $10^2$ NOT and $\sqrt{SWAP}$ operations. Although this number of operations is small compared to that required for the implementation of a quantum algorithm and a quantum error correction code (*e.g.* $N = 10^4$), we envisage that it could be significantly increased, at least by an order of magnitude, by further engineering the composition and structure of the QDs, for example by using isotopically purified (*e.g.* II-VI semiconductors)[16] and/or by deuteration of the QD capping ligands.[39] Furthermore, exciton-Mn spin entangled states could be manipulated by exploiting Raman coherences with sub-picosecond laser pulses,[24] which are at least 3 order of magnitude shorter than the coherence times of Mn spin *qudits* in PbS QDs.

## 3. Conclusions

In conclusion, we have reported on the experimental realization of NOT and $\sqrt{\text{SWAP}}$ universal quantum gates based on the creation and coherent manipulation of $Mn^{2+}$ spin *qudits* in an ensemble of QDs. We have demonstrated full detection of the NMR spectrum of the $Mn^{2+}$ ions, which consists of well defined, addressable spin states that are robust against decoherence phenomena. We have implemented nutation-EDNMR methods for coherent manipulation of hyperfine states and observed Rabi oscillations between selected energy level pairs, enabling the experimental realization of universal quantum gates. Thus, single paramagnetic $Mn^{2+}$ ions confined in solution processable QDs represent a new model spin-*qudit* system beyond the traditional solid state spin *qubits* such as weakly coupled double QDs.[5a, 5b] These findings open up new directions in quantum computation by offering opportunities for integration of QDs in scalable quantum circuits from low-cost, flexible and solution-based fabrication processes.

## 4. Experimental section

*Synthesis of the materials and samples preparation*. The Mn-doped PbS QDs were synthesized in aqueous solution with a Mn weight content of 0.05%, which corresponds to a nominal average number of Mn ions per QD of 0.5. The surface of the QDs was passivated with thiolglycerol (TGL) molecules enabling the realization of nanostructures that are stable and optically active.[21, 40] The core of the QDs has the cubic rocksalt crystal structure of PbS and an average diameter $d = 4.5 \pm 1.2$ nm (**Figure 1a**). For ESR and EDNMR studies, the QD colloidal solutions were freeze dried overnight and inserted into 3 mm outer diameter quartz tubes. The tubes were flushed with nitrogen gas to remove moisture and oxygen contamination and closed with stop cocks. Samples were precooled in liquid nitrogen before insertion in the ESR resonator kept at $T = 5$ K.

*Electron spin resonance.* Pulsed and continuous-wave (CW) ESR experiments were performed on a Q-band ($\nu_{mw} = 34$ GHz) Bruker ElexSys E580 spectrometer coupled to a dielectric resonator at $T = 5$ K. CW-ESR spectra were recorded with magnetic field modulation amplitude and frequency of 0.1 mT and 100 kHz, respectively. CPMG sequence parameters: $\pi = 48$ ns, $\tau = 200$ ns, n (number of $\pi$ pulses) = 200, number of shots per point = 5, shot repetition time = 1 ms. EDNMR sequence parameters: ELDOR pulse length = 0 - 1000 ns, ELDOR frequency width = 626 MHz, $\pi/2 = 24$ ns, $T = 200$ ns, magnetic field $B_0 = 1213 \pm 0.1$ mT.

**Acknowledgements**

This work was supported by the University of Nottingham; The Leverhulme Trust [grant number RPG-2013-242]; the Engineering and Physical Sciences Research Council [grant numbers NS/A000014/1 and EP/K503800/1]; and the Nanoscale and Microscale Research Centre (nmRC) at the University of Nottingham, which provided access to instrumentation for TEM studies. We thank A. Brookfield for technical assistance during the ESR measurements.

**Competing interests**

The authors declare no competing interests

**Figures**

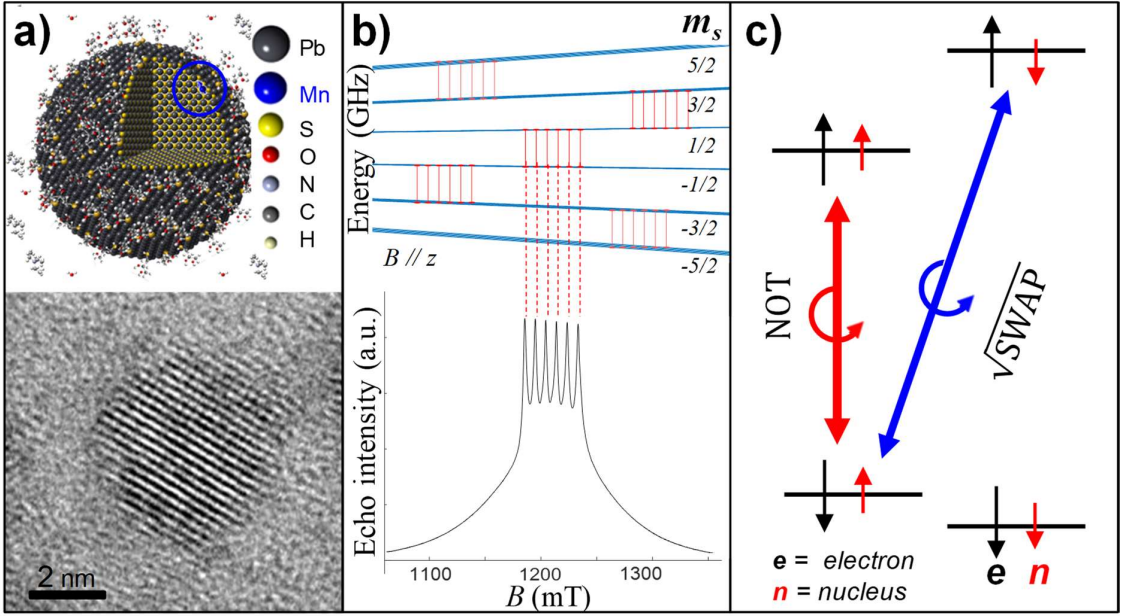

**Figure 1. a)** Schematic representation of a PbS quantum dot containing a Mn-atom and a high resolution TEM image of a representative quantum dot. **b)** Energy level diagram for a Mn spin system with $S = 5/2$ and $I = 5/2$ calculated at Q-band (34 GHz) for $B_0 // z$, $A_{iso} = 267$ MHz, $g = 2$, and $a = 24B_4^4 = 1050$ MHz. along with the full echo detected field swept spectrum for Mn ions in PbS QDs. All the hyperfine transitions occurring between the different $m_s$ states are indicated for a given magnetic field orientation $B_0 // z$. **c)** Realization of logic NOT and $\sqrt{\text{SWAP}}$ quantum gates based on hyperfine interaction between electron, $e$, and nuclear, $n$, spins. A superposition of electronic states between selected energy level pairs is generated and probed by electron double resonance detected nuclear magnetic resonance.

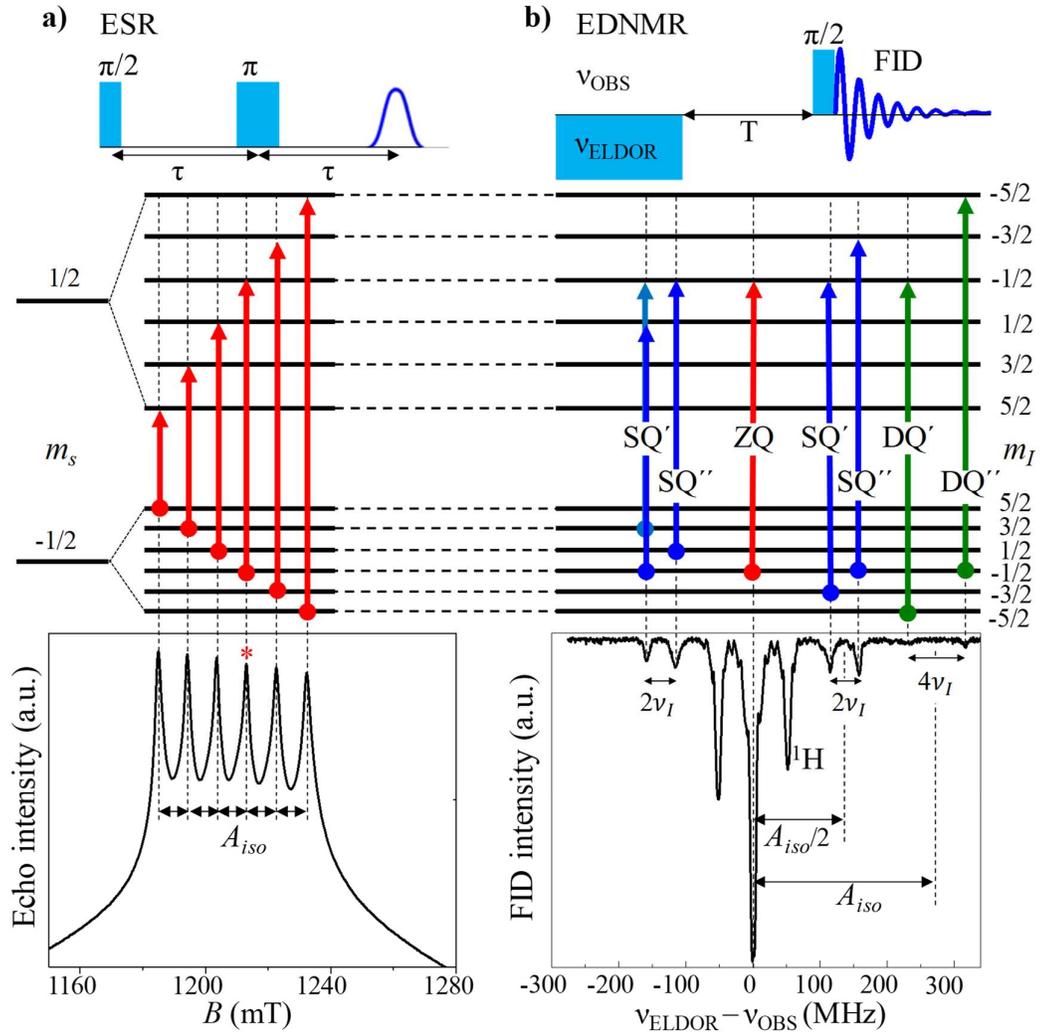

**Figure 2. a)** Electron spin resonance (ESR) and **b)** electron double resonance detected nuclear magnetic resonance (EDNMR) spectra for Mn ions in PbS quantum dots measured at Q-band and $T$ = 5 K. The top-panels show the pulse sequences used to measure the ESR (**a**) and EDNMR (**b**) spectra, and the corresponding energy level diagrams with the ESR and EDNMR transitions. For the sake of clarity the DQ transition connecting the $m_I$ states $|+3/2\rangle$ and $|-1/2\rangle$ and overlapping with the left SQ' transition has been omitted. The spectrum shows the quenching of the free induction decay (FID) when the electron-electron double resonance (ELDOR) frequency, $\nu_{ELDOR}$, matches the observed frequency, $\nu_{obs}$.

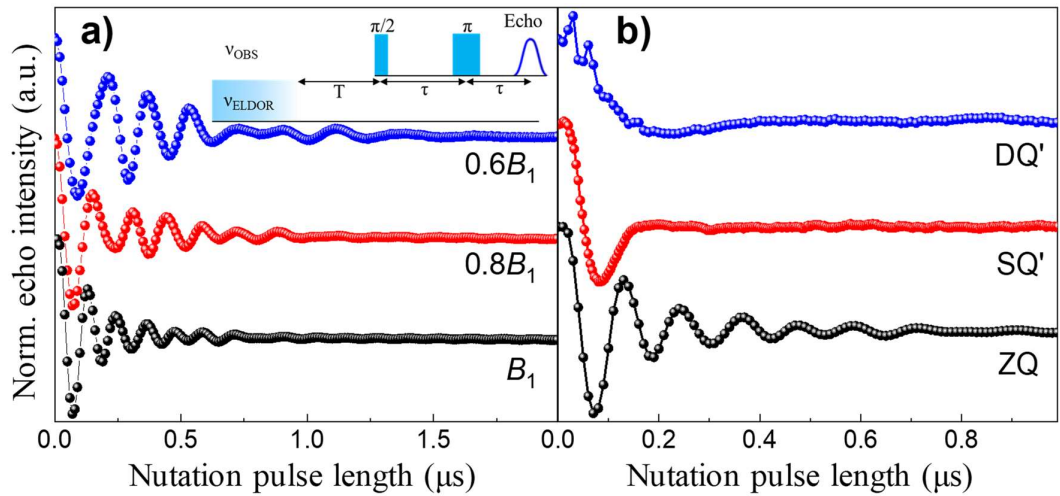

**Figure 3 a)** Nutation-EDNMR data for the ZQ with microwave power dependence. Inset: nutation-EDNMR pulse sequence. **b)** comparison between nutation-EDNMR data for the ZQ, SQ' and DQ' transitions at $B_1 = B_{1max}$ and $T = 5$ K after subtraction of a non-oscillating contribution for SQ´ and DQ´.

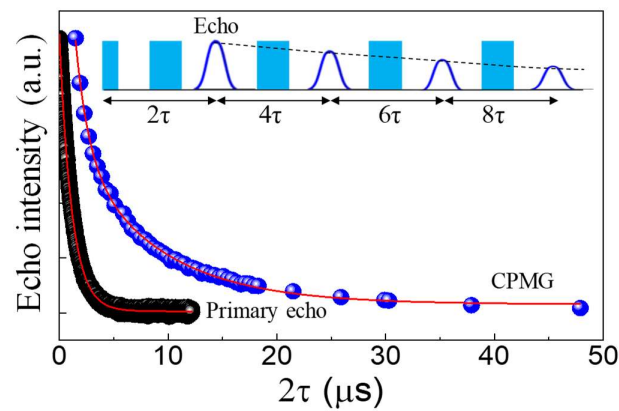

**Figure 4** Comparison between the results for the primary spin echo sequence and the CPMG sequence (inset) recorded at $T = 5$ K and $B = 1221$ mT. Continuous lines are mono and bi-exponential fits to the primary spin echo and CPMG data, respectively (see text).

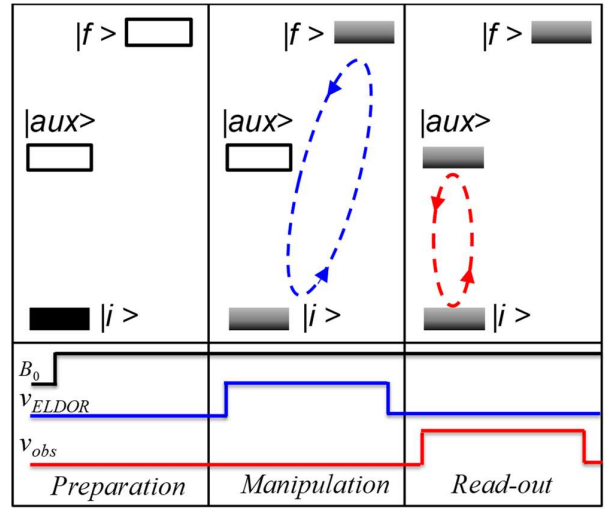

**Figure 5.** Representation of the execution of a quantum gate with a cycle of operations involving preparation, manipulation and read-out (see text). The thickness and colour of $|i>$), $|f>$ and $|aux>$ indicate populations, while arrows indicate population transfer.

# SUPPLEMENTARY INFORMATION

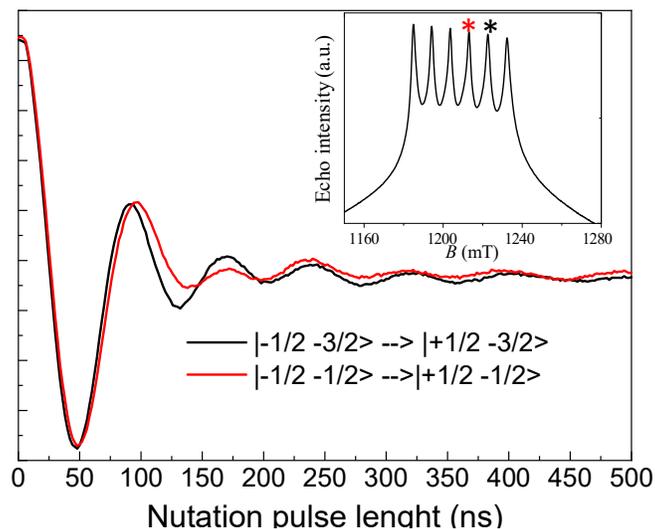

**Figure S1** Nutation-ELDOR data for the ZQ transitions corresponding to the hyperfine lines with $m_I = \pm \frac{1}{2}$ and $m_I = \pm 3/2$. Inset: Echo field swept spectrum measured at T = 5 K. The * symbols indicate the ZQ transitions excited to perform the nutation ZQ oscillations.

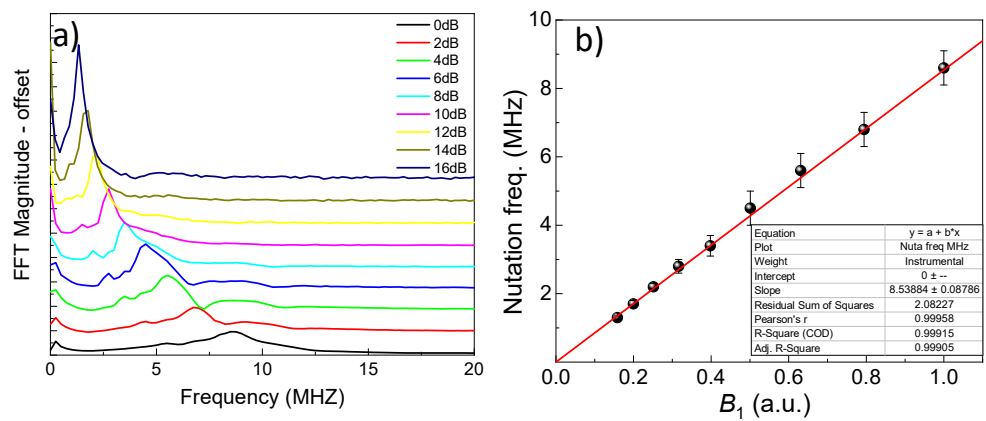

**Figure S2 a)** FFT of the nutation-ELDOR data for the ZQ transition at different microwave attenuation level. **b)** Plot of the dominant frequencies in (a) versus the microwave field $B_1$.

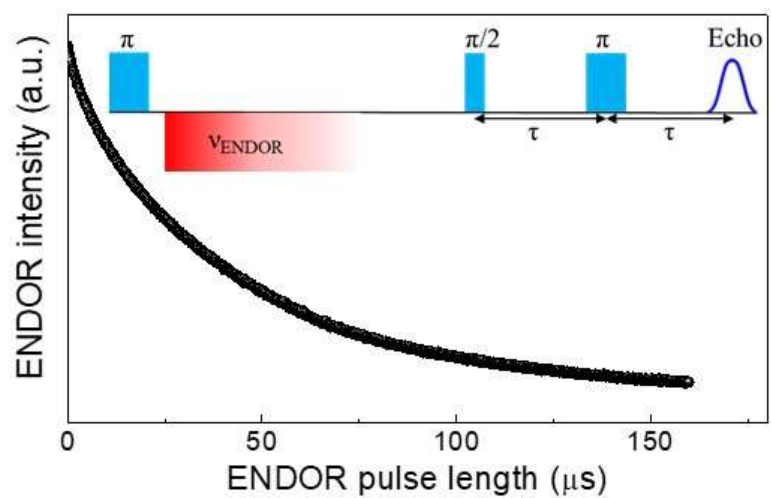

**Figure S3** Nutation-ENDOR data at $T$ = 5 K for the ZQ transition with ENDOR frequency 40 MHz corresponding to the Mn - $^1$H hyperfine transition. Inset, pulse scheme for the implementation of the ENDOR nutation experiment.